\documentclass[
twocolumn,
prl,showpacs,preprintnumbers,amsmath,amssymb]{revtex4}

\usepackage{graphicx}
\usepackage{amssymb}
\usepackage{amsmath}
\usepackage{dsfont}
\usepackage{bm}
\usepackage{mathrsfs}

\begin{document}
\title{Optical Conformal Mapping and Dielectric Invisibility Devices}
\author{Ulf Leonhardt}
\affiliation{
School of Physics and Astronomy, University of St Andrews,
North Haugh, St Andrews KY16 9SS, Scotland
}
\begin{abstract}
An invisibility device should guide light around an object 
as if nothing were there, regardless where the light comes from.
Ideal invisibility devices are impossible 
due to the wave nature of light.
This paper develops a general recipe for the design of media
that create perfect invisibility 
within the accuracy of geometrical optics.
Here the imperfections of invisibility can be made
arbitrarily small to hide objects 
that are much larger than the wavelength.
Using modern metamaterials, 
practical demonstrations of such devices seem possible.
The method developed here can be also applied 
to escape from detection 
by other forms of waves such as sound.
\end{abstract}
\date{\today}
\pacs{42.15.-i, 02.40.Tt}
\maketitle

According to Fermat's Principle \cite{BornWolf},
light rays take the shortest optical paths in dielectric media.
Here the refractive index $n$
integrated along the ray trajectory defines the path length.
When $n$ is spatially varying the shortest optical paths are
not straight lines, but are curved, in general.
This light bending is the cause of many optical illusions.
For example, in a mirage in the desert \cite{Feynman},
light rays from the sky
are bent above the hot sand where the air is thin
and the refractive index is low, because in this way
the rays minimize their optical paths,
creating images of the sky that deceive the observer
as illusions of water \cite{Feynman}.
Imagine a different situation where 
a medium guides light around a hole in it.
Suppose that all parallel bundles of incident 
rays are bent around the hole 
and recombined in precisely the same direction 
as they entered the medium.
An observer would not see the difference between
light passing through the medium
or propagating across empty space 
(or, equivalently, in a uniform medium).
Any object placed in the hole would be hidden from sight.
The medium would create 
the ultimate optical illusion: invisibility \cite{Gbur}.

However, Nachman \cite{Nachman} and 
Wolf and Habashy \cite{WolfHabashy}
proved that perfect invisibility is unachievable,
except in a finite set of discrete directions
where the object appears to be squashed
to infinite thinness
and for certain objects that are small
compared with the wavelength 
\cite{Kerker}. 
In order to carry images, though, light should 
propagate with a continuous range 
of spatial Fourier components,
{\it i.e.} in a range of directions.
The mathematical reason for the impossibility 
of perfect invisibility is the uniqueness of the 
inverse-scattering problem for waves \cite{Nachman}:
the scattering data,
{\it i.e.} the directions and amplitudes of the transmitted 
plane-wave components determine the spatial profile
of the refractive index \cite{Nachman}.
Therefore, the scattering data of light in empty space
are only consistent with the propagation through empty space.
Perfect illusions are impossible due to the wave nature of light.

On the other hand, Nachman's theorem \cite{Nachman}
does not limit the imperfections of invisibility,
they may be very small, 
nor does the theorem apply to light rays, {\it i.e.} 
to light propagation within the regime of geometrical optics
 \cite{BornWolf}.
Here we develop a general recipe, accompanied by an example,
for the design of media that create perfect invisibility
for light rays over a continuous range of directions.
Since this method is based on geometrical optics \cite{BornWolf}, 
the inevitable imperfections of invisibility can be made
exponentially small for objects that are much larger 
than the wavelength of light.

To manufacture a dielectric invisibility device, 
media are needed 
that possess a wide range of the refractive index
in the spectral domain where the device should operate.
In particular, 
Fermat's Principle \cite{BornWolf} seems to imply that
$n<1$ in some spatial regions,
because only in this case 
the shortest optical paths may go around the object
without causing phase distortions.
In our example, $n$ varies from $0$ to about $35.9$.
In practice, one could probably accept a certain degree 
of visibility that significantly reduces the demands 
on the range of the refractive index.

Extreme values of $n$ occur when the material is
close to resonance with the electromagnetic field.
Metamaterials  \cite{Smith}
with man-made resonances can be manufactured
using appropriately designed circuit boards, 
similar to the ones used for demonstrating negative refraction
\cite{Shelby}.
In this research area, 
the quest for the perfect lens  \cite{Pendry}
has lead to spectacular recent improvements
\cite{Smith,Recent,Grigorenko}
mainly focused on the magnetic susceptibilities so far.
In such metamaterials, 
each individual circuit plays the role of an artificial atom
with tunable resonances.
With these artificial dielectrics,
invisibility could be reached for frequencies
in the microwave to terahertz range. 
In contrast, stealth technology is designed 
to make objects of military interest as black as possible to radar. 
There, using impedance matching \cite{Jackson}, 
electromagnetic waves are absorbed without reflection, 
{\it i.e.} without any echo detectable by radar.
Recently, nanofabricated metamaterials with 
custom-made plasmon resonances have been demonstrated
\cite{Grigorenko} that operate in the visible range
of the spectrum 
and may be modified to reach invisibility here.

Our method is also applicable to
other forms of wave propagation, 
for example to sound waves,
where the index $n$ 
describes the ratio of the local phase velocity 
of the wave to the bulk value,
or to quantum-mechanical matter waves
where external potentials act like refractive-index
profiles \cite{BornWolf}. 
For instance,
one could use the profiles of $n$ described here
to protect an enclosed space
from any form of sonic tomography.
But, for having a definite example in mind, 
we focus on light in media throughout this paper.
We study the simplest non-trivial case of invisibility,
an effectively two-dimensional problem.

Consider a dielectric medium that is
uniform in one direction and light of wavenumber $k$
that propagates orthogonal to that direction. 
The medium is characterized by the refractive-index profile $n(x,y)$.
In order to satisfy the validity condition of geometrical optics,
$n(x,y)$ must not significantly vary 
over the scale of an optical wavelength $2\pi/k$
\cite{BornWolf}.
To describe the spatial coordinates in the propagation plane we 
use complex numbers $z=x+iy$
with the partial derivatives
$\partial_x=\partial_z+\partial_z^*$ and
$\partial_y=i\partial_z-i\partial_z^*$
where the star symbolizes complex conjugation.
In the case of a gradually varying refractive-index profile
both amplitudes $\psi$ of the two polarizations of light 
obey the Helmholtz equation \cite{BornWolf}
\begin{equation}
\left(4\partial_z^*\partial_z + n^2 k^2\right) \psi = 0 \,,
\label{eq:helmholtz}
\end{equation}
written here in complex notation with the Laplace operator
$\partial_x^2+\partial_y^2 = 4\partial_z^*\partial_z$.
Suppose we introduce new coordinates $w$ 
described by an analytic function $w(z)$
that does not depend on $z^*$.
Such functions define conformal maps \cite{Nehari} 
that preserve the angles between the coordinate lines.
Since $\partial_z^*\partial_z = |dw/dz|^2 \partial_w^*\partial_w$,
we obtain in $w$ space
a Helmholtz equation with the transformed refractive-index profile $n'$
that is related to the original one as
\begin{equation}
n=n'\left|\frac{dw}{dz}\right| \,.
\label{eq:n}
\end{equation}
Suppose that the medium is designed such that $n(z)$
is the modulus of an analytic function $g(z)$.
The integral of $g(z)$ defines a map $w(z)$
to new coordinates where, according to Eq.\ (\ref{eq:n}),
the transformed index $n'$ is unity. 
Consequently, in $w$ coordinates
the wave propagation
is indistinguishable from empty space
where light rays propagate along straight lines.
The medium performs an
{\it optical conformal mapping} to empty space.
If $w(z)$ approaches $z$ for $w\rightarrow\infty$
all incident waves appear at infinity as if they have travelled through
empty space, regardless what has happened in the medium.
However, as a consequence of the Riemann Mapping Theorem
\cite{Nehari} nontrivial $w$ coordinates occupy Riemann sheets
with several $\infty$, one on each sheet.
Consider, for example, the simple map 
\begin{equation}
w = z + \frac{a^2}{z} \,,\quad
z = \frac{1}{2}\left(w\pm\sqrt{w^2-4a^2}\right)  \,,
\label{eq:simple}
\end{equation}
illustrated in Fig.\ 1,
that is realized by the refractive-index profile $n=|1-a^2/z^2|$. 
\begin{figure}[h]
\begin{center}
\vspace*{-2mm}
\includegraphics[width=18.0pc]{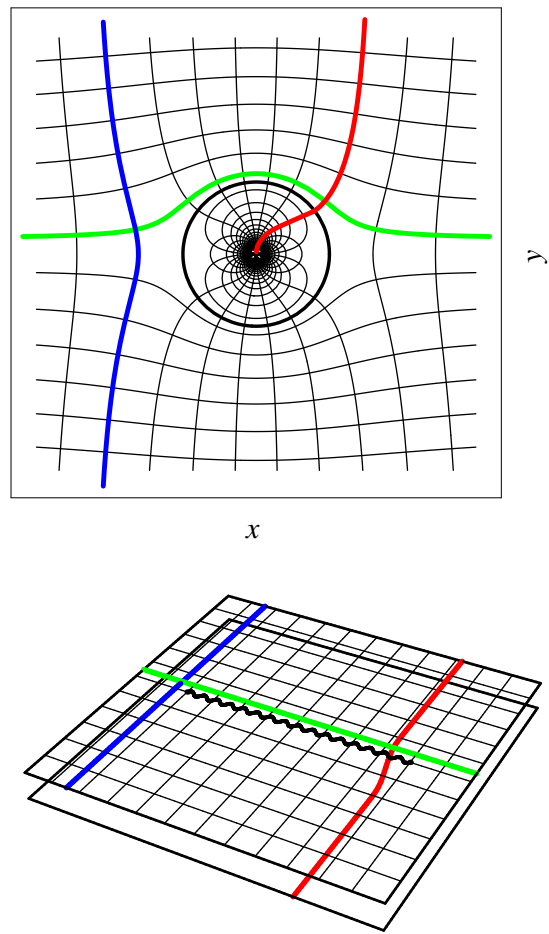}
\vspace*{-13mm}
\caption{
\tiny{
Optical conformal map.
A dielectric medium conformally maps physical space
described by the points $z=x+iy$ of the complex plane
onto Riemann sheets if the refractive-index profile
is $|dw/dz|$ with some analytic function $w(z)$. 
The figure illustrates the simple map (\ref{eq:simple})
where the exterior of a circle in the picture above is 
transformed into
the upper sheet in the picture below, and
the interior of the circle is mapped onto the lower sheet.
The curved coordinate grid of the upper picture is the 
inverse map $z(w)$ of the $w$ coordinates, 
approaching a straight rectangular grid at infinity.
As a feature of conformal maps, the right angles between
the coordinate lines are preserved.
The circle line in the figure above
corresponds to the branch cut between the sheets below
indicated by the curly black line.
The figure also illustrates the typical fates of light rays
in such media.
On the $w$ sheets rays propagate along straight lines.
The rays shown in blue and green avoid the branch cut
and hence the interior of the device.
The ray shown in red crosses the cut and passes onto the 
lower sheet where it approaches $\infty$.  
However, this $\infty$ corresponds to a singularity
of the refractive index and not to the $\infty$ of
physical space. 
Rays like this one would be absorbed, 
unless they are guided back to the exterior sheet.}}
\end{center}
\vspace*{-8mm}
\end{figure}
The constant $a$ characterizes the spatial extension of the medium.
The function (\ref{eq:simple}) maps the exterior 
of a circle of radius $a$ on the $z$ plane onto
one Riemann sheet and the interior onto another.
Light rays traveling on the exterior $w$ sheet may have 
the misfortune of passing the branch cut
between the two branch points $\pm 2a$.
In continuing their propagation, 
the rays approach $\infty$ on the interior $w$ sheet.
Seen on the physical $z$-plane, 
they cross the circle of radius $a$ and 
approach the singularity of the refractive index at the origin. 
For general $w(z)$, only one $\infty$ on the Riemann structure
in $w$ space corresponds to the true $\infty$ of physical $z$ space
and the others to singularities of $w(z)$. 
Instead of traversing space, light rays may cross the branch cut
to another Riemann sheet where they approach $\infty$.
Seen in physical space, the rays are irresistibly attracted towards
some singularities of the refractive index. 
Instead of becoming invisible, the medium casts a shadow
that is as wide as the apparent size of the branch cut is.
Nevertheless, the optics on Riemann sheets turns out
to serve as a powerful theoretical tool for developing the design
of dielectric invisibility devices.

All we need to achieve is to guide light back from the interior 
to the exterior sheet, {\it i.e.}, seen in physical space,
from the exterior to the interior layer of the device.
To find the required refractive-index profile,
we interpret the Helmholtz equation in $w$ space
as the Schr\"odinger equation \cite{BornWolf}
of a quantum particle of effective mass $k^2$ 
moving in the potential $U$ with energy $E$ 
such that $U-E=-n'^2/2$ \cite{BornWolf}.
We wish to send all rays that have passed through the branch cut
onto the interior sheet back to the cut at precisely the same 
location and in the same direction they entered.
This implies that we need a potential 
for which all trajectories are closed. 
Assuming radial symmetry for $U(w)$ around one
branch point $w_1$, say $+2a$ in our example,
only two potentials have this property,
the harmonic oscillator and the Kepler potential \cite{LL1}.
In both cases the trajectories are ellipses \cite{LL1}
that are related to each other by a transmutation of force
according to the Arnol'd-Kasner theorem
\cite{ArnoldKasner}.
The harmonic oscillator corresponds to the  
transformed refractive-index profile $n'$ with
\begin{equation}
n'^2 = 1 - \frac{|w-w_1|^2}{r^2}
\label{eq:ho}
\end{equation}
where $r$ is a constant radius.
The Kepler potential with negative energy $E$ 
is realized by the profile with
\begin{equation}
n'^2 = \frac{r}{|w-w_1|}-1 \,.
\label{eq:kepler}
\end{equation}
Note that the singularity of the Kepler profile in $w$ space
is compensated by the zero of $|dw/dz|$ at a branch point
in physical space such that the total refractive index (\ref{eq:n})
is never singular.
In both cases (\ref{eq:ho}) and (\ref{eq:kepler}), 
$r$ defines the radius of the circle on the interior $w$ sheet
beyond which $n'^2$ would be negative and hence inaccessible
to light propagation. 
This circle should be large enough to cover the branch cut.
The inverse map $z(w)$ turns the outside of the circle into the
inside of a region bounded by the image $z(w)$ 
of the circle line in $w$ space.
No light can enter this region. 
Everything inside is invisible.

Yet there is one more complication:
light is refracted \cite{BornWolf} 
at the boundary between the exterior and the interior layer.
Seen in $w$ space, light rays encounter here a transition
from the refractive index $1$ to $n'$.
Fortunately, refraction is reversible.
After the cycles on the interior sheets light rays are refracted 
back to their original directions, 
as illustrated in Fig.\ 2. 
\begin{figure}[h]
\begin{center}
\includegraphics[width=16.0pc]{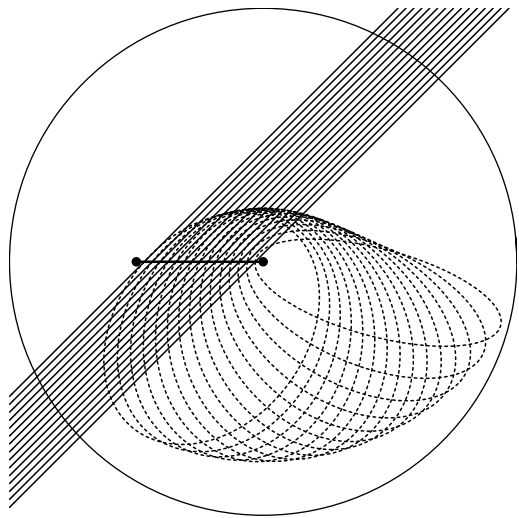}
\vspace*{-1mm}
\caption{
\tiny{
Light guiding.
The device guides light that has entered its interior layer
back to the exterior,
represented here using two Riemann sheets that
correspond to the two layers, seen from above.
At the branch cut, the thick line between
the two points in the figure, the branch points,
light passes from the exterior to the interior sheet.
Here light is refracted according to Snell's law.
On the lower sheet, the refractive-index profile
(\ref{eq:kepler}) guides the rays to the exterior sheet
in elliptic orbits with one branch point as focal point.
Finally, the rays are refracted back to their original directions
and leave on the exterior sheet as if nothing has happened. 
The circle in the figure indicates 
the maximal elongations of the ellipses.
This circle limits the region in the
interior of the device that light does not enter. 
The outside of the circle corresponds to the inside of the device.
Anything beyond this circle is invisible.
}}
\end{center}
\vspace*{-8mm}
\end{figure}
The invisibility is not affected, unless the rays are totally reflected.
According to Snell's Law \cite{BornWolf},
discovered by Ibn Sahl more than a millennium ago \cite{Rashed},
rays with angles of incidence $\theta$ with respect
to the branch cut enter the lower sheet with angles $\theta'$
such that $n'\sin\theta'=\sin\theta$.
If $n'<1$ this equation may not have real solutions
for $\theta$ larger than a critical angle $\Theta$.
Instead of entering the interior layer of the device 
the light is totally reflected \cite{BornWolf}.
The angle $\Theta$ defines the acceptance angle of the
dielectric invisibility device, 
because beyond $\Theta$ the device appears silvery 
instead of invisible.  
The transformed refractive-index profiles 
(\ref{eq:ho}) and (\ref{eq:kepler})
at the boundary between the layers are lowest 
at the other branch point $w_2$ that limits the branch cut,
$w_2=-2a$ in our example.
In the case of the harmonic-oscillator profile (\ref{eq:ho})
$n'$ lies always below $1$ and we obtain the acceptance angle
\begin{equation}
\Theta = \arccos\left(\frac{|w_2-w_1|}{r}\right) \,.
\label{eq:accept}
\end{equation}
For all-round invisibility, 
the radius $r$ should approach infinity,
which implies that the entire interior sheet is employed 
for guiding the light back to the exterior layer.
Fortunately, the Kepler profile  (\ref{eq:kepler}) 
does not lead to total reflection if $r \ge 2|w_2-w_1|$. 
In this case, the invisible area is largest for
\begin{equation}
r = 2|w_2-w_1| \,.
\end{equation}
Figure\ 3 illustrates the light propagation in a dielectric 
invisibility device based on the simple map (\ref{eq:simple})
and the Kepler profile (\ref{eq:kepler}) with $r=8a$.
Here $n$ ranges from $0$ to about $35.9$, 
but this example is probably not the optimal choice.
One can chose from infinitely many conformal maps $w(z)$ 
that possess the required properties for achieving invisibility:
$w(z) \sim z$ for $z\rightarrow\infty$ 
and two branch points $w_1$ and $w_2$. 
The invisible region may be deformed to any
simply-connected domain by 
a conformal map that is the numerical solution of 
a Riemann-Hilbert problem \cite{Ablowitz}.
We can also relax the tacit assumption that $w_1$
connects the exterior to only one interior sheet,
but to $m$ sheets where light rays return after $m$ cycles.
If we construct $w(z)$ as
$af(z/a)$ with some analytic function $f(z)$ 
of the required properties and a constant length scale $a$
the refractive-index profile $|dw/dz|$ is identical
for all scales $a$.
Finding the most practical design is an engineering problem
that depends on practical demands. 
This problem may also inspire
further mathematical research on conformal maps, 
in order to find the optimal design and to extend
our approach to three dimensions. 
\begin{figure}[t]
\begin{center}
\vspace*{-1mm}
\includegraphics[width=18.0pc]{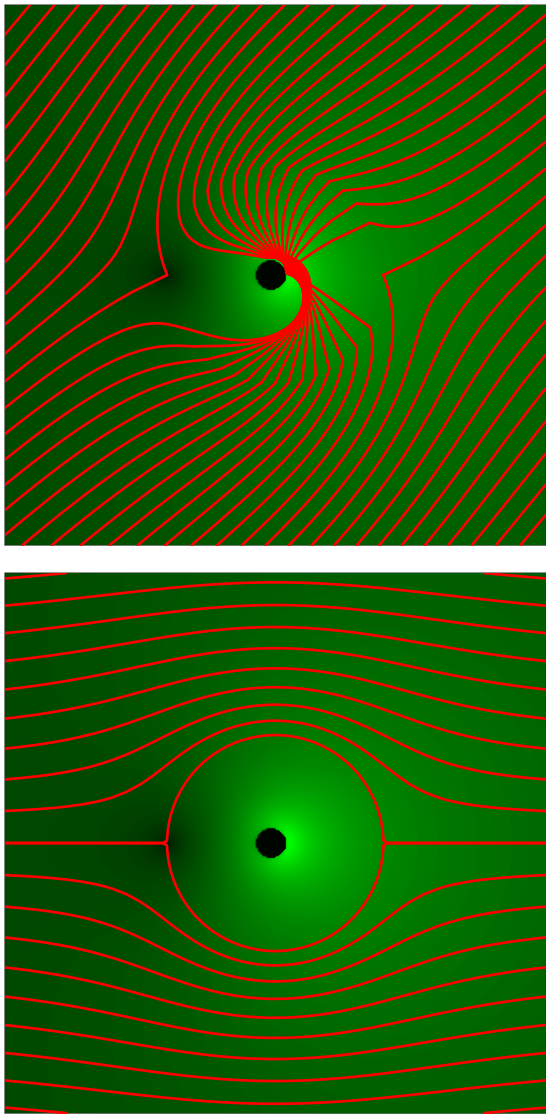}
\vspace*{-6mm}
\caption{
\tiny{
Ray propagation in the dielectric invisibility device.
The light rays are shown in red. 
The brightness of the green background 
indicates the refractive-index profile 
taken from the simple map (\ref{eq:simple})
and the Kepler profile (\ref{eq:kepler}) with $r=8a$
in the interior layer of the device.
The invisible region is shown in black.
The upper figure illustrates how light is refracted at
the boundary between the two layers
and guided around the invisible region
where it leaves the device as it nothing were there.
In the lower figure, 
light simply flows around the interior layer.
}}
\end{center}
\vspace*{-8mm}
\end{figure}

Finally, we return to the starting point and ask why our scheme
does not violate Nachman's theorem \cite{Nachman} 
that perfect invisibility is unattainable.  
The answer is that waves are not only refracted
at the boundary between the exterior and the interior layer,
but also reflected, and that the device causes a time delay.
However, the reflection can be significantly reduced
by making the transition between the layers gradual
over a length scale much larger  than the wavelength $2\pi/k$
or by using anti-reflection coatings. 
In this way the imperfections of invisibility 
can be made as small as the accuracy limit 
of geometrical optics \cite{BornWolf},
{\it i.e.} exponentially small.
One can never completely hide from waves, but from rays.

I am grateful to
Leda Boussiakou,
Luciana Davila-Romero,
Mark Dennis,
Malcolm Dunn,
Greg Gbur,
Clare Gibson,
Julian Henn
and
Awatif Hindi
for the discussions that led to this paper.
My work has been supported by the 
Leverhulme Trust and the
Engineering and Physical Sciences Research Council.

\end{document}